# T. Nadareishvili[1], A. Khelashvili[2]

*(Iv. Javakhishvili Tbilisi State University. Institute of High Energy Physics University str. 9.  0109 Tbilisi.Georgia)*

# GENERALIZATION OF HYPERVIRIAL AND FEYNMAN-HELLMANN THEOREMS FOR SINGULAR POTENTIALS

Using well-known methods we generalize (hyper)virial theorems to case of singular potential. Discussion is performed for most general second order differential equation, which involves all physically interesting cases, as Schrödinger and Klein-Gordon equations with singular potentials. Some physical consequences are discussed. The connection with Feynman-Hellmann like theorems are also considered and some relevant differences are underlined.

## I.     Introduction

Virial theorem has a wide application in classical as well as in quantum mechanics. This theorem connects average values of kinetic and potential energies for the systems confined in limited areas. Moreover it allows making definite conclusions about some interesting problems without solving to equations of motion.

There are many generalizations of virial theorem, especially in relativistic quantum mechanics for investigating of bound states [1].

Recently much attention was devoted to singular potentials, namely, to potentials, behaving as $r^2 V(r) \to -V_0 \;; (V_0 > 0)$ for $r \to 0$ in the Schrodinger equation, and as $rV = -V_0$ for $r \to 0$ in the Klein-Gordon and Dirac equations.

Such behaved potentials appear in large classes of physical problems. Particularly, in Calogero model [2], Coulomb or Hulthen potential in Klein-Gordon and Dirac equations [3], Black Hole theory [4] and etc. Virial like theorems can make things clear while studying such problems.

Therefore, it seems natural to make attempts for generalization of virial theorem for such (singular) potentials too.

The most general methods for obtaining various virial like theorems were developed in [5] by C. Quigg for regular potentials in the Schrodinger equation. The general character of these methods allows us to carry over singular potentials as well. It appears that formally the theorem almost keeps the form familiar for regular potentials with obvious differences.

---

[1] Electronic address:teimuraz.nadareishvili@tsu.ge
[2] Electronic address:anzor.khelashvili@tsu.ge

But the main difference is provided by additional solutions, which are the relevant property of singular potentials and is related to the necessity of self-adjoint extension (SAE).

This article is organized as follows:

First of all we remember the needed methods for deriving of virial like theorems and apply them to general second order differential equation.

Consequences for regular potentials are reviewed and then the singular potentials are considered. It is shown, that there arise additional terms in the usual virial like theorems, which depend on the additional solution for singular potential. Some consequences of the new form of virial theorems are also considered.

After that the corresponding corrections to the Feynman-Hellmann theorem are discussed.

## II. Derivation of Hypervirial (generalized virial) Theorems

Let us consider the second order differential equation of most general form (exclusion of first derivative terms is always possible by using suitable transformation [6])

$$u''(r) + L(r)u(r) = 0, \qquad (2.1)$$

where L(r) is an arbitrary function of r . Central potential in three-dimensions will be important for us in following. Exactly to equation (2.1) reduces the radial Schrodinger equation with $0 < r < \infty$. Even the one-dimensional case may be investigated on the same foot, as well, where $-\infty < x < \infty$. In the following some of physical requirements will be used to restrict this function, L(r).

Now we proceed to the methods of C.Quigg [5]. Let us multiply (2.1) by $fu'$ and integrate in the interval $(0,\infty)$. (here $f(r)$ is an arbitrary three-times differentiable function, which will be restricted somewhere in the following). We derive

$$-\int_0^\infty fu'u''dr = \int_0^\infty fLuu'dr \qquad (2.2)$$

Let us mention that using the following relations $u'u'' = \frac{1}{2}(u')^2{}'$ and $uu' = \frac{1}{2}(u^2)'$, one can perform partial integration in (2.2)

$$-f[u']^2\Big|_0^\infty + \int_0^\infty f'[u']^2 dr = fLu^2\Big|_0^\infty - \int_0^\infty f'Lu^2 dr - \int_0^\infty fL'u^2 dr \qquad (2.3)$$

For bound states $u, u' \to 0$ at large distances and therefore one neglects contributions from upper bound in (2.3), if $f$ and $L$ are to be restricted as follows

$$\lim_{r\to\infty} fu'^2 \to 0; \qquad \lim_{r\to\infty} fLu^2 \to 0 \qquad (2.4)$$

(For scattering problems $u, u'$ are not decaying functions and the conditions (2.4) may not take place, if we do not require it by special choice of $f$).

Therefore there remain expressions in (2.3) only at lower bound

$$fu'^2\Big|_0 + \int_0^\infty f'u'^2 dr = -fLu^2\Big|_0 - \langle f'L\rangle - \langle fL'\rangle, \qquad (2.5)$$

where $<\ >$ denotes averaging by means of $u$ function. For example,

$$\langle fL \rangle = \int_0^\infty fLu^2 dr \qquad (2.6)$$

Now perform a partial integration in the second term of RHS of eq. (2.5), using evident relation $(uu')' = u'u' + uu''$. It follows

$$I \equiv \int_0^\infty f'u'^2 dr = f'uu' \Big|_0^\infty - \int_0^\infty f'uu'' dr - \int_0^\infty f''uu' dr \qquad (2.7)$$

For bound states the first term on RHS at the upper limit may be neglected, if

$$\lim_{r \to \infty} f'uu' \to 0 \qquad (2.8)$$

Now let us integrate the last term on RHS of (2.7)

$$I_1 = \int_0^\infty f''uu' dr = \frac{1}{2} \int_0^\infty f''(u^2)' dr = \frac{1}{2} f''u^2 \Big|_0^\infty - \frac{1}{2} < f''' > \qquad (2.9)$$

For bound states $f$ must be restricted as follows

$$\lim_{r \to \infty} f''u^2 \to 0 \qquad (2.10)$$

Therefore, we have (accounting equation of motion (2.1))

$$I = -f'uu' \Big|_0 + < fL > + \frac{1}{2} f''u^2 \Big|_0 + \frac{1}{2} < f''' > \qquad (2.11)$$

At last, from (2.5) and (2.11) for bound states we derive the following hypervirial theorem:

$$\left\{ fu'^2 - f'uu' + \frac{1}{2} f''u^2 - fuu'' \right\}_{r=0} = -2 < fL > - < fL' > - \frac{1}{2} < f''' > \qquad (2.12)$$

For scattering states (2.4), (2.8) and (2.10) restrictions are not satisfied and instead of (2.12) we have

$$-\left\{ fu'^2 - f'uu' + \frac{1}{2} f''u^2 - fuu'' \right\}_{r=\infty} + \left\{ fu'^2 - f'uu' + \frac{1}{2} f''u^2 - fuu'' \right\}_{r=0} =$$

$$= -2 < fL > - < fL' > - \frac{1}{2} < f''' > \qquad (2.13)$$

After substitution here $u$ function at infinity corresponding hypervirial theorem can be derived for scattering problems as well.

Now let us make some comments in connection to (2.12) about restrictions on $f$:

(a) Because $<\ >$ means averaging by $u$-functions, $f$ must be such, that corresponding integrals do exist.
(b) When $f = r^q$ $(q \geq -2l)$, then (2.12) coincides with (2.27) from [7], in which only the Schrödinger equation is considered, i.e.

$$L = 2m \left[ E - V - \frac{l(l+1)}{2mr^2} \right] \qquad (2.14)$$

with regular V.

Let us note that the choice $f = r^q$ satisfies to (2.4), (2.8) and (2.10) restrictions.

(c) (2.12) - like expression for arbitrary $f$ is derived in [8], but here also, as in [7], only Schrodinger equation was considered.

## III. Some Applications of Hypervirial Theorem

Choosing $f$, one can obtain several interesting expressions from (2.12). Let us consider some of them.

Consider a particular case for $L(r)$ in (2.1)

$$L = A(r) - \frac{s(s+1)}{r^2}, \quad s \geq 0 \qquad (3.1)$$

i.e. we separate a centrifugal term.

We use here a general notation $A(r)$ instead of (2.14) because a lot of physical equations reduce to form, like (3.1), where potential participates in different manners.

It is necessary to make distinction between two cases: $\lim_{r \to 0} r^2 A(r) = 0$ (regular) and $\lim_{r \to 0} r^2 A(r) \neq 0$ (singular).

Consider each of them in detail:

(i) **regular case**, when

$$\lim_{r \to 0} r^2 A(r) = 0 \qquad (3.2)$$

It is easy to guess, that only regular potentials

$$\lim_{r \to 0} r^2 V(r) = 0 \qquad (3.3)$$

obey to (3.2) in case of Schrödinger equation (if we take $s = l; \quad l = 0,1,2$)

While, for example, for one- and two-particle Klein-Gordon equations (3.2) is satisfied if

$$\lim_{r \to 0} rV(r) = 0 \qquad (3.4)$$

When (3.2) is satisfied it follows the following behavior of wave function at the origin

$$u \underset{r \to 0}{\approx} a_s r^{s+1} + b_s r^{-s} \qquad (3.5)$$

The second term in (3.5) does not obey to condition of hermiticity for Hamiltonian [9,10] and radial momentum operator $p_r = -i(\frac{\partial}{\partial r} + \frac{1}{r})$ [11], which is imposed on the wave function at origin

$$\lim_{r \to 0} rR(r) = \lim_{r \to 0} u(r) = u(0) = 0 \qquad (3.6)$$

Therefore it is neglected as a rule (see, any textbook in quantum mechanics). Then at small distances only the first term remains

$$u_s \approx a_s r^{s+1} \qquad (3.7)$$

Substituting this into (2.12) one obtains

$$a_s^2 \left\{ r^{2s} \left[ (s+1)f - (s+1)f'r + \frac{r^2}{2} f'' \right] \right\}_{r=0} = -2 < f'A > - < fA' > + \\ + 2s(s+1) < \frac{f'}{r^2} - \frac{f}{r^3} > - \frac{1}{2} < f''' > \qquad (3.8)$$

Now consider special form for $f$ [5]

$$f = r^q \tag{3.9}$$

We have

$$\left\{(s+1)(1-q) + \frac{1}{2}q(q-1)\right\}a_s^2 r^{q+2s}\Big|_{r=0} = -<2qr^{q-1}A + r^q A'> - \\ -\left[2s(s+1)(1-q) + \frac{1}{2}q(q-1)(q-2)\right]<r^{q-3}> \tag{3.10}$$

In order the LHS of this expression be not diverging at $r = 0$, we must require

$$q \geq -2s \tag{3.11}$$

Therefore, (3.10) becomes

$$(2s+1)^2 a_s^2 \delta_{q,-2s} = \\ = -<2qr^{q-1}A + r^q A' + \left[2s(s+1)(1-q) + \frac{1}{2}q(q-1)(q-2)\right]r^{q-3}> \tag{3.12}$$

It must be noted that (3.12) is a generalization of relation (2.30) from paper [5] for $L$ in form (3.1).

Let now consider various interesting values of $q$ in (3.12):

a) $q = 1$

Then it follows from (3.12) that

$$\langle 2A + rA' \rangle = 0 \tag{3.13}$$

In case of Schrodinger equation, when

$$A = 2m(E - V) \tag{3.14}$$

we derive

$$E = \left\langle V + \frac{1}{2}rV' \right\rangle \tag{3.15}$$

which is the usual virial theorem

$$<T> = \frac{1}{2}\langle rV' \rangle \tag{3.16}$$

b) $q = -2l$

Taking into account separability of total wave function

$$\psi(r,\theta,\varphi) = R_{ne}(r)Y_{em}(\theta,\varphi) = \frac{u_{n,e}(r)}{r}Y_{em}(\theta,\varphi) \tag{3.17}$$

we derive

$$(2l+1)^2 \left|R_{n,e}^{(e)}(0)\right|^2 = (l!)^2 <4l\frac{A}{r^{2l+1}} - \frac{A'}{r^{2l}}>_{n,e} \tag{3.18}$$

Here $R_{n,e}^{(e)}(0)$ is the l-th order derivative of radial wave function at origin. (3.18) generalizes eq. (1.4) of [7] for Schrodinger equation

$$(2l+1)^2 \left|R_{n,e}^{(e)}(0)\right|^2 = 2m(l!)^2 \left[\left\langle \frac{1}{r^l}\frac{dV}{dr}\right\rangle_l + 4l\left\langle \frac{E-V}{r^{2l+1}}\right\rangle\right] \tag{3.19}$$

c) $q=0$ or $f = const$

This case is well-known in the Schrodinger equation [5,8]. Now it follows from (2.12):
$$\{u'^2 - uu''\}_{r=0} = -<L'> \qquad (3.20)$$
or
$$(l+1)a_l^2 r^{2l}\big|_{r=0} = -<A'(r)> - \left\langle \frac{2l(l+1)}{r^3} \right\rangle \qquad (3.21)$$

If now we take $l = 0$, then
$$a_0^2 = (u_0)'^2(0) = -<A'(r)> \qquad (3.22)$$

It generalizes eq. (39a) from [8] to arbitrary $A(r)$. When we take expression (3.14), then it follows from (3.22) the well-known relation
$$|\psi_0(0)|^2 = \frac{m}{2\pi}\left\langle \frac{dV}{dr} \right\rangle \qquad (3.23)$$

In case $l \neq 0$, the LHS of (3.21) is zero and therefore we obtain
$$2l(l+1)\left\langle \frac{1}{r^3} \right\rangle = -\langle A' \rangle \qquad (3.24)$$

which generalizes eq. (39b) from [8] for arbitrary $A(r)$. The relations (3.22) and (3.24) are formulated in terms of $A(r)$. Depending on equations of motion, the potential $V(r)$ appears in various forms and one must take care, which restrictions arise on potential $V(r)$.

d) $q \neq 0, 1, -2l$

Now we have
$$< 2qr^{q-1}A + r^q A' + \left[2l(l+1)(1-q) + \frac{1}{2}q(q-1)(q-2)\right]r^{q-3} >= 0 \qquad (3.25)$$

This expression allows us to connect average values of various degrees of $r$.
For example, in Schrodinger equation we have
$$2Eq\langle r^{q-1} \rangle - 2q\langle r^{q-1}V \rangle - \langle r^q V' \rangle + \frac{(q-1)}{m}\left[\frac{q}{4}(q-2) - l(l+1)\right]\langle r^{q-3} \rangle = 0 \qquad (3.26)$$

For power-like potential, $V = V_0 r^n$ it follows from (3.26), that
$$2Eq\langle r^{q-1} \rangle - V_0(2q+n)\langle r^{q+n-1} \rangle + \frac{(q-1)}{m}\left[\frac{q}{4}(q-2) - l(l+1)\right]\langle r^{q-3} \rangle = 0 \qquad (3.27)$$

If $n = -1$, the well-known Kramer's formula [12] follows for the Coulomb potential $V = -\frac{\alpha}{r}$, (i.e. $V_0 = -\alpha; \; q = s+1$)
$$2E(s+1)\langle r^s \rangle + \alpha(2s+1)\langle r^{s-1} \rangle + \frac{s}{m}\left[\frac{s^2-1}{4} - l(l+1)\right]\langle r^{s-2} \rangle = 0 \qquad (3.28)$$

But when $n = 2$, the relation for isotropic harmonic oscillator $V = \frac{1}{2}\omega^2 r^2$ is derived [13]

$$2E(s+1)\langle r^s \rangle - \omega^2(s+2)\langle r^{s+2} \rangle + \frac{s}{m}\left[\frac{s^2-1}{4} - l(l+1)\right]\langle r^{s-2} \rangle = 0 \qquad (3.29)$$

Also it is possible to derive recurrence like relations between different powers of $r$ for various relativistic equations. Such relations have many applications in various physical problems [14].

ii) **Singular case.** Now

$$\lim_{r \to 0} r^2 A(r) = -V_0 ; (V_0 > 0) \qquad (3.30)$$

As was shown in [15-16], for Schrodinger and two equal mass particles' Klein-Gordon equations, that besides the standard levels there appear additional levels as well, whose wave function behaves at small distances as

$$u_{st} \approx a_{st} r^{\frac{1}{2}+P} \; ; \qquad u_{add} \approx a_{add} r^{\frac{1}{2}-P} \qquad (3.31)$$

where, for example, in Schrodinger equation

$$P = \sqrt{(l+1/2)^2 - 2mV_0} > 0 \qquad (3.32)$$

while in the Klein-Gordon equation for two equal mass particles

$$P = \sqrt{(l+1/2)^2 - V_0^2/4} > 0 \qquad (3.33)$$

Likely it is possible to find P for given L for each relativistic equation. At the same time, as is indicated in [15-16], for the existence of additional levels following constraint must be satisfied

$$0 \leq P < 1/2 \qquad (3.34)$$

which is expression of vanishing of the radial wave function $u(r)$ at the origin, $u(0) = 0$.

Now if we take the wave function at small distances as general form [15]

$$u = a_{st} r^{\frac{1}{2}+P} + a_{add} r^{\frac{1}{2}-P} \qquad (3.35)$$

and use (3.9) for $f$, then (2.13) gives

$$(1-q)(1/2+P-q/2)a_{st}^2 \delta_{q,1-2P} + (1-q)(1/2-P-q/2)a_{add}^2 \delta_{q,1+2P} + \left[(q-1)^2 - 4P^2\right]a_{st}a_{add}\delta_{q,1} =$$
$$= -\left\langle 2qr^{q-1}A + r^q A' + \left[2l(l+1)(1-q) + \frac{q}{2}(q-1)(q-2)\right]\langle r^{q-3}\rangle\right.$$

(3.36)

Here we must require that $q \geq 1-2P$. If $V_0 = 0$, or we return to regular case (3.2), because the RHS of (3.36) remains unchanged, but the LHS transforms into the LHS of (3.12)

Let consider various q-s in (3.36) as above.

a) $q = 1; \; P \neq 0, 0 < P < 1/2$
Then from (3.36) follows

$$\langle 2A + rA' \rangle = 4P^2 a_{st} a_{add} \qquad (3.37)$$

For the Schrodinger equation this means

$$E = \left\langle V + \frac{1}{2}rV' \right\rangle + \frac{P^2}{m}a_{st}a_{add} \tag{3.38}$$

Therefore, for singular potential the virial theorem differs from that of regular ones by the extra term

$$b = \frac{P^2}{m}a_{st}a_{add} \tag{3.39}$$

This term vanishes when we take only standard or only additional solutions.

***Comment***: Separate consideration needs the case $P = 0$. As is indicated in [15], we have in this case

$$u \underset{r\to 0}{\approx} a_{st} r^{\frac{1}{2}} + a_{add} r^{\frac{1}{2}} \ln r \tag{3.40}$$

Clearly $u(0) = 0$. Now instead of (3.36) it follows

$$\left\langle 2qr^{q-1}A + r^q A' \right\rangle - \left[2l(l+1)(1-q) + \frac{1}{2}q(q-1)(q-2)\right]\left\langle r^{q-3}\right\rangle = 0 \tag{3.41}$$

And virial theorem for Schrodinger theory takes the form

$$E = \left\langle V + \frac{1}{2}rV' \right\rangle \tag{3.42}$$

which is analogous to regular potential case, but difference appears in averaging by function (3.40).

For pure singular potential

$$V = -\frac{V_0}{r^2}; (V_0 > 0) \tag{3.43}$$

it follows from (3.37) that

$$E = \frac{P^2}{m}a_{st}a_{add} \tag{3.44}$$

This is a single level, which appears in quantum mechanical consideration, when we retain the additional solution as necessary ingredient for providing a self-adjointness of Hamiltonian via self-adjoint extension (SAE) [15].

This level disappears immediately as we neglect pure standard or pure additional solutions.

It is evident that the equality (3.37) is rather general relation and many physical consequences can be derived from it.

Consider, for example, two-particle Klein-Gordon equation with equal masses $m$:

$$u'' + \left[\frac{V^2}{4} - \frac{MV}{2} + \frac{M^2}{4} - m^2\right]u - \frac{l(l+1)}{r^2}u = 0; \tag{3.45}$$

$M$ is a total mass of composite state.. Comparison to (2.1) and (3.1) gives

$$A = \frac{V^2}{4} - \frac{MV}{2} + \frac{M^2}{4} - m^2$$

Using this in (3.36), we obtain

$$\left\langle \frac{V^2}{2} - MV + \frac{rV'}{2}(V - M) + \frac{M^2}{2} - 2m^2 \right\rangle = 4P^2 a_{st}a_{add} \tag{3.46}$$

Let us now consider the following problem: Can two massive particles produce massless bound state in case of Coulomb potential (attraction or repulsion)? Existence of bound states for both cases is a consequence of the relativistic structure of Klein-Gordon equation, where for $M = 0$ there remains only $V^2$ in (3.45). This problem was considered in scientific literature [17].

For this aim one must take $M = 0$ in (3.46). We derive

$$\left\langle \frac{V^2}{2} + \frac{rV'}{2}V - 2m^2 \right\rangle = 4P^2 a_{st} a_{add} \qquad (3.47)$$

For Coulomb potential it follows

$$-m^2 = 2P^2 a_{st} a_{add} \qquad (3.48)$$

and we see that the positive answer this problem has only if $a_{st} \neq 0$ and $a_{add} \neq 0$ (if $a_{st} a_{add} < 0$). Correctness of this result may be verified also by direct solution of the Klein-Gordon equation. Indeed, substituting $M = 0$ in (3.45), one finds

$$u'' + \left[\frac{V^2}{4} - m^2\right]u - \frac{l(l+1)}{r^2}u = 0; \qquad (3.49)$$

If we take here $V = \mp \frac{\alpha}{r}$ this equation becomes

$$u'' + \left[-m^2 - \frac{P^2 - 1/4}{r^2}\right]u = 0 \qquad (3.50)$$

where $P$ is given by (3.33). Note that this equation coincides to the Schrodinger equation with the accuracy of notations. Therefore we can use the results of our paper [15] and write down the general solution derived there

$$u(r) = \sqrt{mr}\{AI_P(mr) + BI_{-P}(mr)\} \qquad (3.51)$$

where $I_P$ and $I_{-P}$ are the modified Bessel functions. We have the following behaviour at infinity

$$u(r) \underset{r \to \infty}{\approx} \frac{1}{\sqrt{2\pi}}\{A + B\}e^{mr} \qquad (3.52)$$

Requiring vanishing of $u(r)$ at infinity as for bound state solution we have to take

$$B = -A \qquad (3.53)$$

Remembering the well-known relation

$$K_P(z) = \frac{\pi}{2\sin P\pi}[I_{-P}(z) - I_P(z)] \qquad (3.54)$$

our wave function takes the form

$$u = -A\frac{2}{\pi}\sqrt{mr}\sin P\pi \cdot K_P(mr) \qquad (3.55)$$

which is exponentially damping at infinity and in the interval $0 \leq P < 1/2$ satisfies to fundamental requirement (3.6). It is evident, that our solution is derived by the requirements

$$A \neq 0; \quad B \neq 0 \qquad (3.56)$$

which means, that $M = 0$ state can be derived only by SAE procedure. We see that explicit solution of Klein-Gordon equation repeats the conclusion, derived by Virial theorem.

One important remark is in order: W. Krolikowski [17] derived the same solution for $l = 0$ state only. It is true, because $K_P(z)$ is the only Bessel function, which behaves in a needed fashion at infinity (vanishes!). It appears that a massless bound state for Coulomb potential may be constructed from 2 massive particle in nonzero orbital momentum states also, $l \neq 0$ [15]. But SAE procedure is necessary.

Owing to the fact, that repulsive case also forms a massless bound state, we conclude, that the following <u>alternatives</u> take place:

(i) Those values of SAE parameter $\tau = \dfrac{a_{add}}{a_{st}}$, when this strange fact occurs, must be deflected in order to suppress such unphysical results.

(ii) We must recognize, that the SAE procedure produces an *effective attraction*, which may be seen from equation (3.50), where the factor $(P^2 - 1/4)$ has a negative sign in area (3.34) and gives a quantum anticentrifugal potential, which is attractive [15].

(iii) It is not excepted that such unphysical fact is a pathology of the Klein-Gordon equation. For example if we reverse the problem and ask if two massless particles can compose a massive bound state in Coulomb field, we can easily see that (3.46) gives a positive answer in case of Coulomb repulsion, but not for attraction.

b) Cases $q = 1 \mp 2P$ and $q \neq 0, 1, -2l$ may be discussed in full analogy. One derives some recurrence like relations between average values of various powers of $r$.

## IV. Connection with the Feynman and Hellmann Theorem

It is known that the Feynman-Hellmann (FH) [18-19] and generalized FH [20-21] theorems are closely related to the hypervirial theorems.

FH like theorems connect average values of energy derivative by some parameters to those of Hamiltonians.

We want to take attention to the fact, that for singular potentials in Schrödinger equation, when SAE is necessary, FH theorem also should be modified.

Indeed, in a traditional way [20], we consider a wave equation of the form
$$F(E, \lambda)\psi = 0 \tag{4.1}$$
where $\psi$ is the wave function of a bound state with energy E, which depends on the parameter $\lambda$. Then we can write
$$\frac{\partial}{\partial \lambda}\langle \psi | F | \psi \rangle = \left\langle \frac{\partial \psi}{\partial \lambda} \bigg| F \bigg| \psi \right\rangle + \left\langle \psi \bigg| \frac{\partial F}{\partial E}\frac{\partial E}{\partial \lambda} + \frac{\partial F}{\partial \lambda} \bigg| \psi \right\rangle + \left\langle \psi \bigg| F \bigg| \frac{\partial \psi}{\partial \lambda} \right\rangle = 0 \tag{4.2}$$

If F has the property that (this property is fulfilled in the regular (3.2) case!)
$$\left\langle \psi \bigg| F \bigg| \frac{\partial \psi}{\partial \lambda} \right\rangle = \left\langle F\psi \bigg| \frac{\partial \psi}{\partial \lambda} \right\rangle \tag{4.3}$$

then in view of (4.1), from (4.2) we obtain
$$\frac{\partial E}{\partial \lambda} = -\frac{\left\langle \psi \bigg| \frac{\partial F}{\partial \lambda} \bigg| \psi \right\rangle}{\left\langle \psi \bigg| \frac{\partial F}{\partial E} \bigg| \psi \right\rangle} \tag{4.4}$$

If $F = H - E$, then (4.4) reduces to the usual Feynman – Hellmann theorem [20-21]

$$\frac{\partial E}{\partial \lambda} = \left\langle \psi \left| \frac{\partial \hat{H}}{\partial \lambda} \right| \psi \right\rangle \tag{4.5}$$

Now from (2.1) and (3.1) we have

$$F = \frac{d}{dr^2} + A(r) - \frac{l(l+1)}{r^2} \tag{4.6}$$

and in the singular (3.30) case, instead of (4.3) we obtain additional term on the right side

$$\left\langle u_n \left| F \right| \frac{\partial u_n}{\partial \lambda} \right\rangle = \left\langle F u_n \left| \frac{\partial u_n}{\partial \lambda} \right\rangle + \lim_{r \to 0} \left[ u_n(r,\lambda) \frac{d}{dr} \frac{\partial u_n(r,\lambda)}{\partial \lambda} - \frac{\partial u_n(r,\lambda)}{\partial \lambda} \frac{du_n(r,\lambda)}{dr} \right] \tag{4.7}$$

and it follows

$$\frac{\partial E}{\partial \lambda} = -\frac{\left\langle \psi \left| \frac{\partial F}{\partial \lambda} \right| \psi \right\rangle + B}{\left\langle \psi \left| \frac{\partial F}{\partial E} \right| \psi \right\rangle} \tag{4.8}$$

where

$$B = \lim_{r \to 0} \left[ u_n(r,\lambda) \frac{d}{dr} \frac{\partial u_n(r,\lambda)}{\partial \lambda} - \frac{\partial u_n(r,\lambda)}{\partial \lambda} \frac{du_n(r,\lambda)}{dr} \right] \tag{4.9}$$

We see that FH theorem is modified as well.

This happens because $F$ is not a self-adjoint operator in singular case (3.30) and the fundamental relation

$$\langle \psi | F | \varphi \rangle = \langle F \psi | \varphi \rangle \tag{4.10}$$

does not takes place. Therefore the SAE procedure is necessary.
Inserting (3.35) we obtain

$$B = -2P \left[ a_{n,st} \frac{\partial a_{n,add}}{\partial \lambda} - a_{n,add} \frac{\partial a_{n,st}}{\partial \lambda} \right] + \frac{dP}{d\lambda} \lim_{r \to 0} \left\{ 4P a_{n,st} a_{n,add} \ln r - a_{n,add}^2 r^{-2P} \right\} \tag{4.11}$$

Consider some consequences for Schrodinger and one body Klein-Gordon equation.
a) For Schrodinger equation P is given by (3.32) and for singular

$$\lim_{r \to 0} r^2 V = -V_0 \, (V_0 > 0) \tag{4.12}$$

potential from (4.7) we obtain

$$\frac{\partial E_n}{\partial \lambda} = \left\langle u_n \left| \frac{\partial \hat{H}_r}{\partial \lambda} \right| u_n \right\rangle + \frac{P}{m}\left[ a_{n,st} \frac{\partial a_{n,add}}{\partial \lambda} - a_{n,add} \frac{\partial a_{n,st}}{\partial \lambda} \right] -$$
$$- \frac{1}{2m}\frac{dP}{d\lambda} \lim_{r \to 0}\left\{ 4P a_{n,st} a_{n,add} \ln r - a_{n,add}^2 r^{-2P} \right\}$$
(4.13)

We see that the last parenthesis of eq. (4.13) is divergent expression at the origin, except the regular case or when $a_{add} = 0$. Therefore only for $\frac{\partial P}{\partial \lambda} = 0$ has this expression a viable sense, i.e. when we choose $\lambda \neq m, V_0$ or $l$.

So when SAE procedure (which is necessary in a singular potential case) is not used, the FH theorem takes usual form (4.5).

When P does not depend on $\lambda$ ($\lambda \neq m, V_0$ or $l$), there remains only the first row in (4.13)

$$\frac{\partial E_n}{\partial \lambda} = \left\langle u_n \left| \frac{\partial \hat{H}_r}{\partial \lambda} \right| u_n \right\rangle + \frac{P}{m}\left[ a_{n,st} \frac{\partial a_{n,add}}{\partial \lambda} - a_{n,add} \frac{\partial a_{n,st}}{\partial \lambda} \right]$$
(4.14)

In particular case, when $P = 0$, calculations must be performed by function (3.40). In this case singularities from (4.13) disappear. We find

$$\frac{\partial E_n}{\partial \lambda} = \left\langle u_n \left| \frac{\partial \hat{H}_r}{\partial \lambda} \right| u_n \right\rangle + \frac{1}{2m}\left[ a_{n,st} \frac{\partial a_{n,add}}{\partial \lambda} - a_{n,add} \frac{\partial a_{n,st}}{\partial \lambda} \right]$$
(4.15)

b) For one body Klein - Gordon equation

$$u'' + \left[ (E-V)^2 - m^2 - \frac{l(l+1)}{r^2} \right] u = 0$$
(4.16)

We obtain for $\lambda \neq V_0$ or $l$

$$\frac{\partial E}{\partial \lambda} = \frac{1}{E - \langle V \rangle}\left\{ \left\langle (E-V)\frac{\partial V}{\partial \lambda} \right\rangle - \frac{B}{2} \right\}$$
(4.17)

And for $\lambda = m$ we get

$$\frac{\partial E}{\partial m} = \frac{m}{E - \langle V \rangle} + P\left\{ a_{st} \frac{\partial a_{add}}{\partial m} - a_{add} \frac{\partial a_{st}}{\partial m} \right\}$$
(4.18)

Where

$$P = \sqrt{(l+1/2)^2 - V_0^2} > 0$$
(4.19)

The main result here is that in case of singular potentials Feynman-Hellmann theorem has to be modified.

## V. Conclusions

In this article we consider problems, related to the singular potentials in light of hypervirial and FH theorems. Main results can be summarized as follows:
1. We have derived a hypervirial theorem for general second order differential equation.
2. For regular potentials we generalized known results concerning the Schrodinger equation ( virial theorem, wave function and its derivatives at origin, recurrence relations between average values of different powers of $r$ )
3. We obtain virial theorem for singular potential, by means of which some physical results are derived (existence of one level for pure $r^{-2}$ potential, possibility of having massless bound state for repulsive and attractive Coulomb potential in the two-body Klein – Gordon equation).
4. Modification of Feynman-Hellmann theorem for singular potential, when the SAE procedure is applied.

**ACKNOWLEDGMENTS.**

The authors thank to T.Kereselidze, A.Kvinikhidze M.Nioradze and participants of seminars at Iv. Javakhishvili Tbilisi State University, for many valuable comments and discussions. The designated project has been fulfilled by financial support of the Georgian National Science Foundation (Grant № GNSF/ST07/4-196).

**References**

1. W.Lucha. Mod.Phys.Lett. "Relativistic virial theorems". A5:2473-2484 (1990).
2. B.Basu-Mallick et all. "Quantization and conformal properties of a generalized Calogero model". Eur.Phys.J.C49:875-889, (2007).
3. H.Saad. "The Klein-Gordon equation with a generalized Hulthen potential in D-dimensions". math- ph/0709.4014(2007).
4. K.Maharana. "Symmetries of the near horizon of a Black Hole by Group Theoretic methods"Int.J.Mod.Phys.A 22, 1717(2007).
5. C. Quigg, J.L.Rosner. "Quantum mechanics with applications to quarkonium" Phys.Report.56:167-235,1979.
6. E.Kamke. "Handbook of ordinary differential equations"."Nayka" (1971).(In Russian)
7. A.AKhare. "The relative magnitudes of the derivatives of 1P(D,F,…) and 2P(D,F,…) Wave functions at the origin". Nucl.Phys.B181:347,1981.
8.H.Grosse,A.Martin. "Exact results on potential models for quarkonium systems".Phys.Report.60:341,1980.
9. Pauli W 1958 *Die allgemeinen Prinzipen der Wellenmechanik* In *Handbuch der Physik*,Bd.5,vol(Berlin: Aufl)
10. D.I.Blokhincev. "Foundations of quantum mechanics". "Nauka".(1976)

11. A.Messhia. "Quantum mechanics".Vol 1; Wiley. New York.(1966)
12. H.M.Kramers."Quantum Mechanics".North-Holland Publishing Company,Amsterdam (1954).
13. J.N.Epstein and S.T.Epstein. "Some Applications of Hypervirial Theorems to the Calculations of Average Values".Am.J.Phys.30,266(1962)
14. G.S.Adkins."Dirac-Coulomb energy levels and expectation values". Am.J.Phys.76,579(2008).
15. A.A.Khelashvili, T.P.Nadareishvili. Bulletin of Georgian Acad.Sci.Vol 164 N1 (2001).
16. T. Nadareishvili, A. Khelashvili. arXiv:0903.0234.math-ph(2009)].
17. W.Krolikowski; "Massless Coulomb Bound State Of A Klein-Gordon Pair".Bulletin De Lacademics polonaise.Vol 18,83 (1979).
18. R.F.Feynman. "Forces in Molecules". Phys.Rev. 56, 340 (1939).
19. H.Hellmann.Einfurung in die Quantenchemie ". Deuticke,Leipzig (1937).
20. D.B.Lichtenberg. "Application of a generalized Feynman – Hellmann theorem to bound-state energy levels". Phys.Rev D. 40,4196 (1989).
21. S.Balasubramanian. "A note on the generalized Hellman-Feynman theorem".Am.J.Phys.58,1204 (1990)].